\documentclass[onecolumn,superscriptaddress,secnumarabic,amssymb,amsmath,nobibnotes,
aps,prd,nofootinbib,altaffilletter]{revtex4}
\usepackage{amsmath,amssymb}
\usepackage{graphicx}
\usepackage{dcolumn}
\usepackage{bm,color}
\usepackage{accents}
\usepackage{amssymb,float}
\usepackage{amsmath}
\usepackage{multirow}
\usepackage{siunitx}
\usepackage{tabularx}
\usepackage{booktabs}
\usepackage{url}
\usepackage[normalem]{ulem}
\usepackage{mathrsfs}
\usepackage{adjustbox}

\DeclareSIUnit\parsec{pc} 

\begin{document}

\title[$H_0$ Tension on the Light of Supermassive Black Hole Shadows~Data]{$H_0$ Tension on the Light of Supermassive Black Hole Shadows~Data}

\author{Celia Escamilla-Rivera}
\email{celia.escamilla@nucleares.unam.mx}
\affiliation{Instituto de Ciencias Nucleares, Universidad Nacional Aut\'{o}noma de M\'{e}xico, Circuito Exterior C.U., A.P. 70-543, M\'exico D.F. 04510, M\'{e}xico}

\author{Rub\'en Torres Castillejos}
\email{torres.utx@gmail.com}
\affiliation{Facultad de Ciencias en F\'isica y Matem\'aticas, Universidad Aut\'onoma de Chiapas, Calz. Emiliano Zapata Km. 8, Tuxtla Gutiérrez 29050, Chiapas, Mexico.}


\begin{abstract}
Cosmological tensions in current times have opened a wide door to study new probes to constrain cosmological parameters,  specifically, to determine the value of the Hubble constant $H_0$ through independent techniques. 
The two standard methods to measure/infer $H_0$ rely on: \text{(i)}~anchored observables for the distance ladder, and  \text{(ii)} establishing the relationship of the $H_0$ to the angular size of the sound horizon in the  recombination era assuming a standard Cosmological Constant Cold Dark Matter ($\Lambda$CDM) cosmology. 
However, the former requires a calibration with observables at nearby distances, while the latter is not a direct measurement and is model-dependent. The physics behind these aspects restrains our possibilities in selecting a calibration method that can help  minimise the systematic effects or in considering a fixed cosmological model background. Anticipating the possibility of deeply exploring the physics of new nearby observables such as the recently detected black hole shadows, in this paper we propose standard rules to extend the studies related to these observables. 
Supermassive black hole shadows can be characterised by two parameters: the angular size of the shadow and the black hole mass. We found that it is possible to break the degeneracy between these parameters by forecasting and fixing certain conditions at high(er) redshifts, i.e., instead of considering the $\approx$10\% precision from the EHT array, our results reach a $\approx 4\%$, a precision that could be achievable in experiments in the near future.
Furthermore, we found that our estimations provide a value of $H_0 = 72.89\pm 0.12$ km/s/Mpc and,  for the baryonic mass density, $\Omega_m=0.275\pm 0.002$, showing an improvement in the values  reported so far in the literature.
We anticipate that our results can be a starting point for more serious treatments of the physics behind the SMBH shadow data as cosmological probes to relax tension issues.
\end{abstract}


\maketitle


\section{\label{sec:intro}Introduction}
One of the most challenging problems of modern cosmology is the statistical tension on the estimation of the expansion rate of the universe today: the Hubble constant $H_0$.
Different experiments, observables and predicted theoretical models gives $H_0$ values that disagree strongly. Therefore, in order to determine $H_0$, the distance to observables on astrophysical and cosmological scales is one of the most relevant factors since measurements of this constant in the local universe based on distance ladder methods do not match the estimated value in the early universe, where the first methodology takes into account serious statistical precision analysis and the latter considers a standard cosmological model supported by observational evidence.

The most pressing issue, in particular for this tension, is the 5.0$\sigma$ disagreement between the local value given by the SH0ES collaboration \cite{Riess:2021jrx} ($H_0$ = 73.04 $\pm$ 1.04 km/s/Mpc) and the early estimated value given by the Planck collaboration \cite{Planck:2018nkj} ($H_0$ = 67.27 $\pm$ 0.60 km/s/Mpc). While measurements involved in each sector of the early and late universe agree, e.g., CMB and BAO observations,\footnote{{It is {important} to mention that the combination of CMB lensing and BAO data has raised serious outcomes related to the spatial curvature of the universe, i.e., without considering these observables, Planck data tends to agree with a closed universe scenario~\cite{DiValentino:2019qzk,Handley:2019tkm}.}}  the $H_0$ tension persists at the ends of the empirical distance ladder.   Several proposals have been raised to find a viable solution to this issue. On one hand, it is possible to consider late $H_0$ measurements which do not require a benchmark model, such as the standard Cosmological Constant Cold Dark Matter ($\Lambda$CDM) model. On the other hand, early $H_0$ measurements can be performed if we assume a collection of physical properties based on a pre-established model that can describe the evolution of the universe. A broad compendium of these two paths for estimating $H_0$ can be found in \cite{Abdalla:2022yfr}.

While the core of the latter paths are based on refining our calibration methods or considering physics beyond the standard $\Lambda$CDM model, we should contemplate the possibility of exploring the nature of new observables that could shed some light on the $H_0$ tension. In this direction, astronomical objects with greater diversity, both in distance and physical characteristics, are being used as standard rulers in our local universe.
Among the proposals, it has been suggested that we can use Black Hole (BH) shadows as standard rulers \cite{Tsupko:2019pzg,Vagnozzi:2020quf}.

The detection of the first BH shadow of M87* by the Event Horizon Telescope Collaboration (EHT) \cite{EventHorizonTelescope:2019dse} opened a new window to using this kind of {rulers} to  independently determine the BH mass and its distance from the observer. Through this mechanism, we can compute the physical size of the BH shadow and compare it with the observed size. Along with the  detection of M87* in our nearby galaxy, a second one, the shadow of the central BH in our galaxy,  Sagittarius A* (Sgr A*) \cite{EventHorizonTelescope:2022wkp}, has also been detected. 
 In particular, supermassive black holes (SMBH) shadows are interesting candidates to study our local universe since their physics is quite simple, and we can see them as standard rulers if the relation between the size of the shadow and the mass of the SMBH that produces it, the so-called {angular size redshift} $\alpha$, is established. To use these kinds of observables, we need to set two limits: 
\begin{enumerate}
\item Measurements at low redshifts. SMBH shadows can be used to estimate $H_0$ in a cosmological independent way and also without evoking the distance ladder method. By adopting a peculiar velocity of the host galaxy $v_p$ in km/s, the mass of the SMBH $M$ in solar masses $M_\odot$ and the angular size $\alpha$, we can directly compute  the distance to the SMBH. Finally, using the Hubble law, we can estimate $H_0$. Clearly, performing this kind of estimate using two SMBH shadows' data points (M87* and Sgr A*) is insufficient, not only by the low data point density, but also due their high uncertainties. However, we expect a future improvement in this direction since the abundance of SMBHs can be hosted in spiral and elliptical galaxies \cite{Zubovas2012}.
At this scale,\cite{Renzi:2022fmw} proposed using a mock SMBH shadow catalog as anchor in the distance ladder method, which can offer an estimation of $H_0$.
\item Measurement at high redshifts. At this scale, determining the mass of the BH can be difficult due to the fact that we need high resolution in the equipment, and the estimation of the uncertainties is quite large. However, reverberation mappings~\cite{Shen:2014uby} techniques combined with spectroastrometry analyses \cite{Wang:2019gaq} have been employed to determine BH mass and distance simultaneously. In \cite{Qi:2019zdk}, the authors  proposed a set of simulated SMBH shadows at this scale, which were performed by assuming a fiducial benchmark cosmology, making it possible to determine a set of cosmological parameters as $\Omega_m$ and $H_0$.
\end{enumerate}

These approaches have set a convenient path to estimate $H_0$. It is important to mention that both of them require certain assumptions between a large SMBH shadows baseline to perform the statistics or higher resolutions in the equipment to perform the observations. SMBH shadows as a cosmological probe are still new and lack statistical power in comparison to other baselines. However, the future of these measurements looks bright and hopeful. While the technological capacity is developing, we can start by considering more serious analyses behind their physics. Our goal is to forecast a larger set of SMBH shadows by relaxing the assumptions made in the latter approaches. By doing this, we will significantly  decrease the errors, and we will obtain a higher $H_0$ in comparison to the reported ones. 

Recent works have studied the possibility of using BH shadow measurements from the Event Horizon Telescope (EHT), e.g., to constrain astrophysical free parameters on a Kerr--Newman--Kiselev--Letelier BH configuration \cite{Atamurotov:2022nim}, analyse cosmological constant corrections on the BH shadow radii  \cite{Adler:2022qtb}, determine optical features from a  Schwarzschild MOG BH with several thin accretions \cite{Deng2022}, examine BH charges \cite{Zhang:2022tpr} and evaluate the effects in the Kerr--Newman BH in quintessential dark energy scenarios \cite{Khan:2020ngg}.

This paper is organised as follows. 
In Section~\ref{sec:BH}, we review the characteristics in order to employ SMBH shadows as standard rulers, and we describe the equations behind these. 
In Section~\ref{sec:BHsim}, we present the algorithm to simulate SMBH shadows and compare them with EHT observations. In addition, we include the process to perform this forecasting at low and high(er) redshifts.
In Section~\ref{sec:data}, we describe the current data sets employed in our analyses including the compilation of the BH events and their observations (see Table \ref{tab:EHT data}), and the new forecasted baseline. 
In Section~\ref{sec:results}, we discuss the results obtained for our $H_0$ estimations performed with our mock data and present a comparison with previous works. 
Finally, we give a summary of discussions in Section~\ref{sec:discussion}.


 \section{\label{sec:BH} Black Hole Shadow Description as Standard Rulers}

A general way to describe the BH shadow in a realistic expanding universe is to consider the angular size/redshift relation, which establishes the information between the apparent angular size of the BH and its redshift and how this quantity changes at cosmological distances. Once this angular size is determined, we can compute the angular diameter distance to the BH. The advantage to describing these physical distance relations is inspired by the fact that SMBH shadows can be used as standard rulers \cite{Tsupko:2019pzg}, which makes them useful to determine $H_0$ in two regimes:
\begin{itemize}
\item Nearby galaxies (low redshift, $z\leq 0.01$), where SMBH shadows are cosmologically model-independent 
and do not require methods that consider anchors in the distance ladder. By a simple calculation using the Hubble flow velocity of a galaxy and its peculiar velocity, we can derive the Hubble velocity and constrain $H_0$. In addition, at $z\rightarrow 0$, the angular size of the BH shadow grows as the BH mass gets bigger, which makes it a suitable candidate to study our local universe in comparison to the BAO peaks, in which amplitude decreases due to the cosmological expansion. However, since the angular size and the mass of the BH are linearly correlated, we require precise techniques to break this degeneracy such as, for example, stellar-dynamics~\cite{Gillessen:2008qv}, gas-dynamics \cite{Lupi2015} or maser observations \cite{Pesce:2020xfe,Kuo2010}. 
\item High(er) distance observations (high(er) redshift $0.01< z < 7$ ($z>7$)), where not only do we require a fiducial cosmology to determine the BH mass, but also we require an angular resolution around 0.1 $\mu$as. These kinds of observations have been show to be extremely difficult~\cite{Tsupko:2019pzg}; however, by combining techniques such as reverberation mappings~\cite{Shen:2014uby} and very long baseline interferometry technologies, it will be possible to achieve such resolution. Going further in distance, the search for  quasars beyond $z>7$ \cite{Matsuoka2019} allows us to estimate BH masses at higher redshift. While the uncertainties in these measurements are high, it is important to note that this is a reference that SMBH shadows can be part of future catalogs, making them useful in proving the cosmic expansion at this scale.  
\end{itemize}

As far as we know, there have been two methodologies to constrain $H_0$ using SMBH shadows in the described redshift ranges:
\begin{itemize}
\item Combination of M87$^{*}$ observation and SNIa catalog \cite{Qi:2019zdk}. This proposal allows us to study a particular set of cosmological parameters by considering  a collection of 10 BH shadows' simulated data points under a fiducial benchmark cosmology ($\Omega_m=0.3$ and $H_0 = 70$ km/s/Mpc) and the Pantheon SNIa catalog (see Section~\ref{sec:data} for its description). The results are reported in Table \ref{tab:dataref}. However, these simulations are restricted solely to BH masses of $ M=3 \times 10^9 M_{\odot}$ within an interval $7< z < 9$, which, combined with SNIa observations, are not able to constrain the $\Lambda$CDM model.
\item  Combination of mock catalogues for SMBH ($\approx 10^6$ BH simulated data points) plus mock SNIa data for the Vera C. Rubin Observatory LSST\footnote{\href{https://www.lsst.org}{www.lsst.org}}.  This proposal starts from the same point of view as the latter; however, the mock SMBH data are used as anchors to calibrate the distance ladder. While the number of BH data points simulated is high, the forecasting is based on a benchmark cosmology and a single shadow data from M87$^{*}$. Furthermore, a cosmographic approach was employed at low redshift, making it impossible to constrain atthe  third order of the series, i.e., the jerk current value~$j_0$~\cite{Renzi:2022fmw}. 
\end{itemize}

Our goal is to forecast a larger set of SMBH shadows by relaxing the assumptions made in the latter methodologies. Additionally, we are going to consider a second shadow data: the Sgr A* observation \cite{EventHorizonTelescope:2022wkp}.  

To set the theoretical quantities, we need to write distance quantities in terms of the characteristics of the object under study. In this work, we employ the standard definition of the luminosity distance for a flat $\Lambda$CDM model as \cite{Weinberg}: 
\begin{equation} \label{eq:luminosity}
     d_L(z) = (1+z)\frac{c}{H_0}I(z) ,
\end{equation}
where  $c$ is the speed of light, $H_0$ is the present-day Hubble parameter and  $I(z)$ is given by the integral
\begin{equation} \label{eq:cosmointegral}
 I(z) = \int_{0}^{z} \left(\Omega_{m0}(1+\Tilde{z})^3 + \Omega_{\Lambda_{0}}\right)^{-1/2} ~d\Tilde{z} , 
\end{equation}
where $\Omega_{m0}$ and $\Omega_{\Lambda_{0}}$ are the present values of the critical density parameters for matter and a dark energy component, respectively. 
The luminosity distance can be related to the angular diameter distance $d_A(z)$ by the reciprocity theorem which states that \cite{Etherington2007}:
\begin{equation} \label{eq:reciprocity}
    d_L(z) = (1+z)^2 d_A(z).
\end{equation}
By definition, the angular diameter distance of an object is $d_A = L / \Delta \theta$, with $L$ the proper diameter of the object and $\Delta \theta$ its observed angular diameter.
If we are able to measure one of these distances at certain redshift $z$, then we can obtain information about the cosmological parameters denoted in Equation~(\ref{eq:cosmointegral}).

Certainty, a BH does not emit photons. Moreover, we can observe the light rays that curve around its event horizon and create a ring with a black spot in the center. We call this {the} {shadow of the black hole}. 
It is known that for a Schwarzschild (SH) BH, the angular radius of its shadow is \cite{Bisnovatyi-Kogan:2019wdd}:
\begin{equation} \label{angular shadow}
    \alpha_{sh}(z) = \frac{3\sqrt{3}m}{d_A(z)},
\end{equation}
where $m = GM / c^2$ is called the mass parameter of the black hole,  $G$ being the constant of gravitation, $c$ the speed of light and $M$ the mass of the BH in solar masses units. This equation is an approximate expression for the visible angular radius of the shadow when using the angular diameter distance $d_A$ by assuming a radial coordinate large enough in comparison with the SMBH horizon in order to obtain an effective linear radius as $3\sqrt{3}m$.
Using Equations~(\ref{eq:luminosity})--(\ref{eq:reciprocity}), we can rewrite the above equations as:
\begin{equation}\label{eq:luminosity shadow}
    \alpha_{sh} = \frac{3\sqrt{3}m}{(1+z)} \frac{c}{H_0}I(z).
\end{equation}
Notice that the BH shadow depends on its mass and distance, the Hubble constant $H_0$ and the free cosmological parameters ($\Omega_m, \Omega_\Lambda$).

Using Equations~(\ref{eq:cosmointegral}) and (\ref{eq:luminosity shadow}), notice that at lower redshift we obtain
\begin{equation}\label{eq:lowred}
\alpha_{\text{LR}} = \frac{3\sqrt{3}m H_0}{cz}, 
\end{equation}
where we can easily notice that given a value for the radius of the shadow $\alpha_{LR}$ and the redshift $z$ of the BH with a known mass, it is possible to directly determine a value of $H_0$.

With Equation~(\ref{eq:reciprocity}), we can establish the relation between the modulus distance and the distance modulus of the source through:
\begin{equation} \label{eq:modulus}
     \mu = m(z) - M = 5~ log_{10} ~d_L(z) +25,
\end{equation}
where $m(z)$ and $M$ are the apparent and absolute magnitude of the supernova, respectively. 

Notice that in the latter equations, we have considered that the SMBHs are described by an SH metric with spin zero. However, astrophysical SMBHs have rotation, and the spin effects can change the size described by Equation~(\ref{angular shadow}). This leads us to consider a Kerr metric, where both the SMBH spin and the observer inclination angle will change the BH shadow along the horizontal axis.  Meanwhile, in the vertical axis line of sight, we conserve the SH shadow. This asymmetric shadow size has been observed by the EHT collaboration; nevertheless, a set of numerical phenomenological expressions is required in order to compute the right size of this shadow. A further study on these key characteristics has been conducted in \cite{Renzi:2022fmw}. In the discussion section of this work, we mention how this treatment does not affect the determination of the distance from the shadow.


\section{\label{sec:BHsim}Black Hole Shadows Forecasting}

Detecting BH shadows requires a great deal of time and technological resources. In recent years, the EHT collaboration\footnote{\href{https://eventhorizontelescope.org}{eventhorizontelescope.org}} has been able to observe two SMBH shadows, M87* and Sgr A*, and continues to work in search of more of them and in different systems, e.g., binary systems. However, two data points are not statistically enough to generate comprehensive astrophysical analyses let alone  cosmological ones. In addition, reaching a sufficient number of observations that can significantly constrain cosmological parameters can take a long period of time. Until we can reach an optimal number of observations of such a kind, 
we can perform forecasting through the standard rule methods.

\begin{figure}[H]
    \centering
    \includegraphics[width=9.2cm]{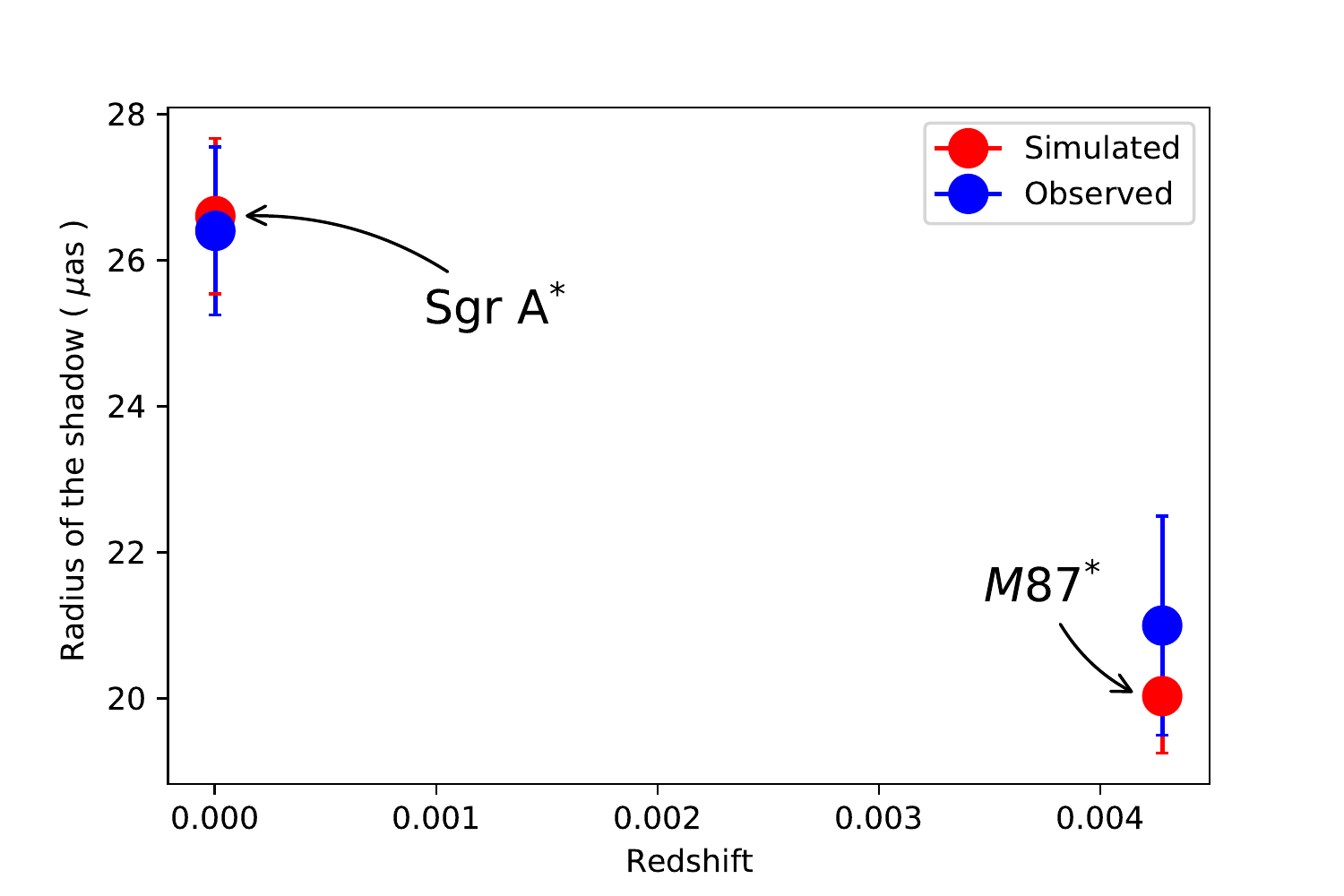}
    \caption{Comparison of the forecasted SMBH shadow radius (red color) from EHT observations  M87* and Sgr A* (blue color). We consider a conservative fixed cosmology with $H_0 = 73.8$ km/s/Mpc, $\Omega_{m0}=0.262$ to compare the observable radius $\alpha$ with its  forecasted value. } 
    \label{fig:comparison}
\end{figure}

In order to produce  mock data for SMBH shadows M87* and Sgr A*, we use \mbox{Equation~(\ref{eq:luminosity shadow})} and its low redshift approximation $\alpha_{LR}$ (\ref{eq:lowred}). 
Following this line of thought, we consider these steps:
\begin{enumerate}
\item  Assume a conservative fiducial cosmological model. In our case, instead of using a conservative Planck data cosmology \cite{Planck:2018vyg} as in other studies, we will consider the local values: $H_0= 73.8$ km/s/Mpc, $\Omega_{m0} = 0.262$, where $\Omega_{\Lambda_{0}} = 1 - \Omega_m$. 

\item Under the above condition, we compute Equation~(\ref{eq:cosmointegral}) only as a function of the redshift $z$. Once the integral is solved, we can use the reciprocity relation Equation~(\ref{eq:reciprocity}) in  Equation~(\ref{eq:luminosity shadow}). This will be our main function, and it takes as input values a set of two variables: the redshift $z$ and the SMBH mass. We can consider this result for the low redshift case as a cutoff when $z \leq 0.1$. Notice that it is necessary to write these equations in BH shadow units, e.g., $M_{\odot}$, and  perform the appropriate conversion to express the results in $\mu$as units. Additionally, we need to assume an error in the simulations. In \cite{Qi:2019zdk}, the authors  considered the M87* single data, which at low redshift constrains $H_0$ with a 13\% error. In order to reduce this number, it was considered a symmetric uncertainty in this single data point as the variance takes the form $P\sqrt{N}=\sigma$; therefore, if we want to reach a precision of $P\approx 4\%$, we require $N=10$ SMBH shadows simulations. While this assumption is reasonable and the estimated error decreases by almost 8\%, the mean value for $H_0$ does not change. Since we are going to consider two SMBH shadows from M87* and Sgr A*, this symmetric uncertainty assumption will be relaxed in order to 
reach a precision of 4\% using a conservative quantity comparable with other observables at low $z$, e.g., SNIa, and a number of simulations derived from data with asymmetric errors at high $z$.

\item In comparison to the previous methodologies described in Section~\ref{sec:BH}, in order to test our algorithm effectiveness, we can compare the simulated SMBH shadows' outputs with the current M87$^{*}$ and Sgr A* observations given by the EHT. As we show in Figure~\ref{fig:comparison}, the simulations are very near to the observables, e.g., for Sgr A* we have that $\alpha_{\text{observed}} = 26.4 \pm 1.15~ \mu$as, while $\alpha_\text{simulated} = 26.6 \pm 1.06~ \mu$as. For the  M87* case, we have a value $\alpha_{\text{observed}} = 21 \pm 1.5~ \mu$as, and our simulation gives $\alpha_\text{simulated} = 19.5 \pm 0.78~ \mu$as. The latter result gives a 4.7\% difference from the simulated data point. This quantity is due to our pre-established precision since the error percentage in the M87* observation is close to 7$\%$, while for Sgr A* it is reduced to almost 4$\%$.
\end{enumerate}

Once we have tested our algorithm against the available observables, we can build a data points larger catalog that can be used in our statistical analyses. To do so, we need to build a baseline with input data containing pairs of redshift and SMBH mass $(z,M)$. In our case, we are going to generate two different mock catalogs: {(i)} for nearby galaxies estimations (low redshift) and {(ii)} for high(er) redshift SMBH shadows. The architecture of our algorithm is shown in Figure~\ref{fig:processb}.


\subsection{\label{subsec:BHsimhigh} High(er) Redshift Observations for Hubble Constant Constraints}

\begin{figure}[H]
    \centering
    \includegraphics[width=10.8cm]{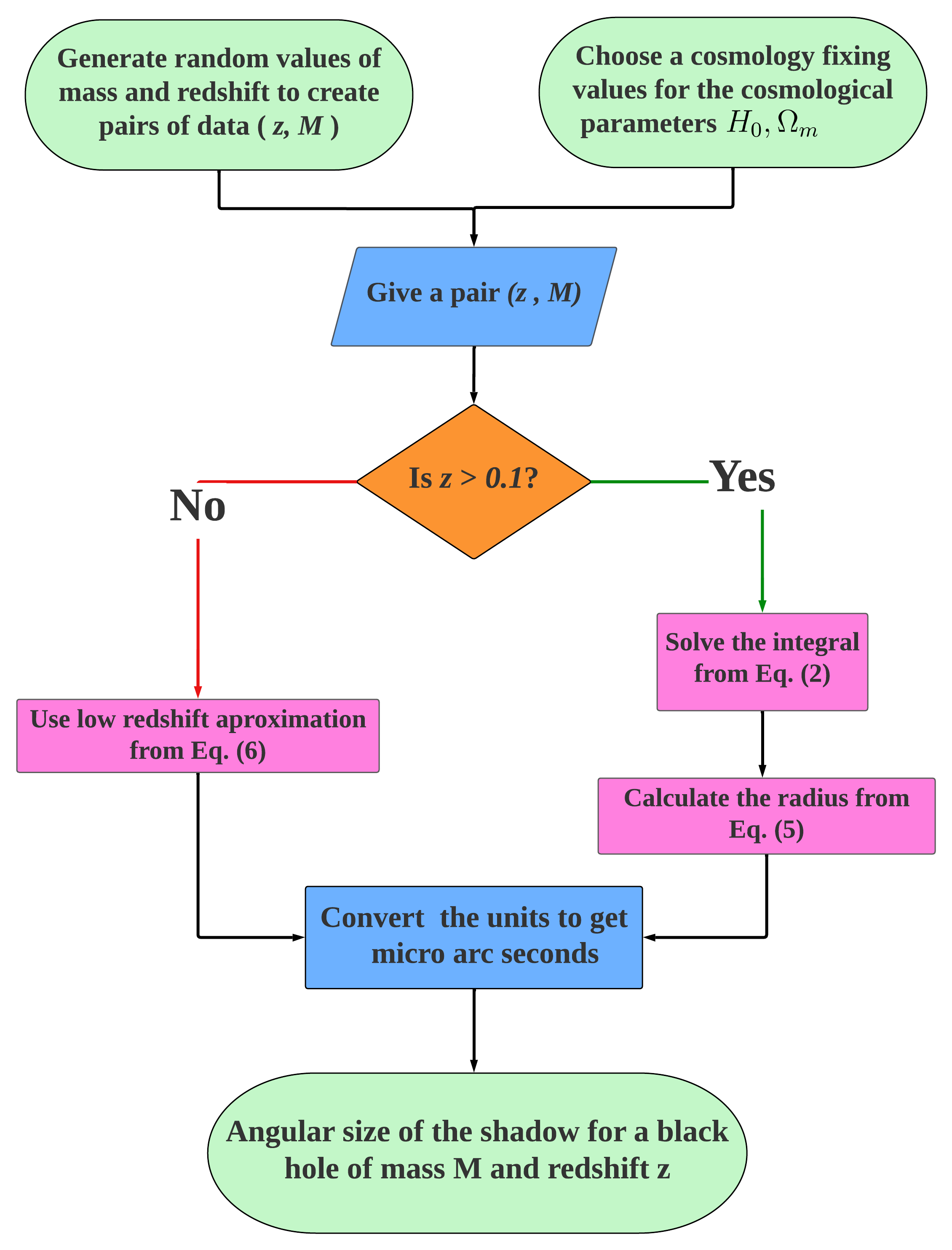}
    \caption{Architecture method to forecast SMBH shadows. The color boxes denote the physical quantities and priors (green color), the setting of variables and units (blue color), the redshift cutoff (orange color) and  the computation employed (pink color). } 
    \label{fig:processb}
\end{figure}

For high(er) redshift, we will simulate SMBH shadows within $7 \leq z \leq 9$. As we mentioned, since some quasar observations can be reached in such a range, it is possible to observe SMBHs in this region if they have a minimum of  $10^9  M_\odot$. These types of BHs are viable as long as they have existed long enough to acquire larger masses, as, e.g., in the case of primordial BH. Although the population of BHs in this range is expected to be large, if we take into account the above conditions, then the number of BHs that meet these characteristics are drastically reduced; therefore, we first generate 10 random redshift values in the range of $7 \leq z \leq 9$ according to Step 2 described above. For the SMBH masses, we will use a uniform random distribution that takes values in the range of $10^9-10^{10}~ M_\odot$. Once we have the synthetic catalog with 10 random pairs as $(z,M)$, we can feed them into the simulation algorithm described in Figure~\ref{fig:processb} and obtain the radii of the SMBH shadow associated with each pair. Our forecasted data is shown at the left of Figure~\ref{fig:simulationsBH}. 

\subsection{\label{subsec:BHsimlow} Nearby Galaxies Estimation Observations for Hubble Constant Constraints}

\begin{figure}[H]
    \centering
    \includegraphics[width=8.5cm]{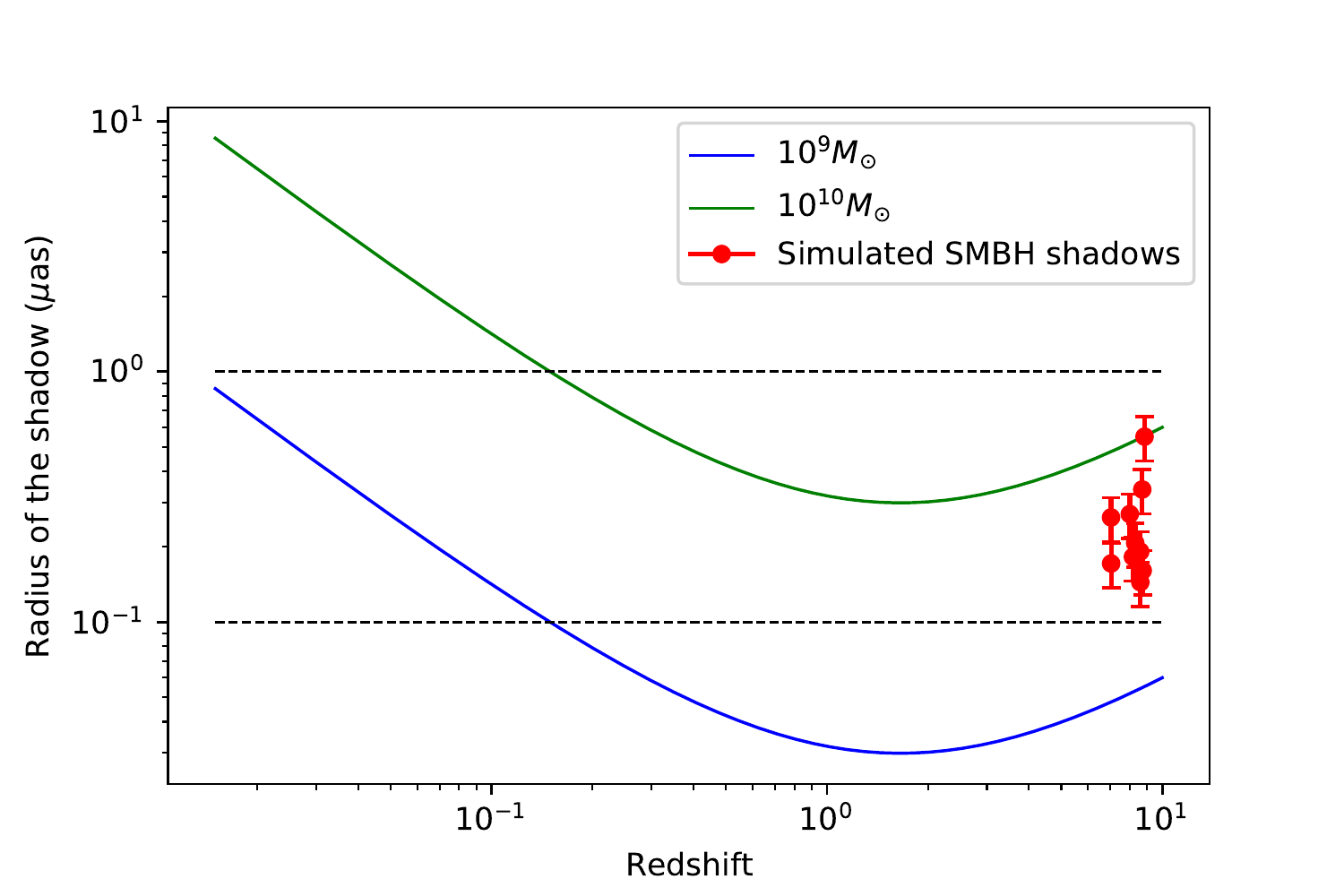}
       \includegraphics[width=8.5cm]{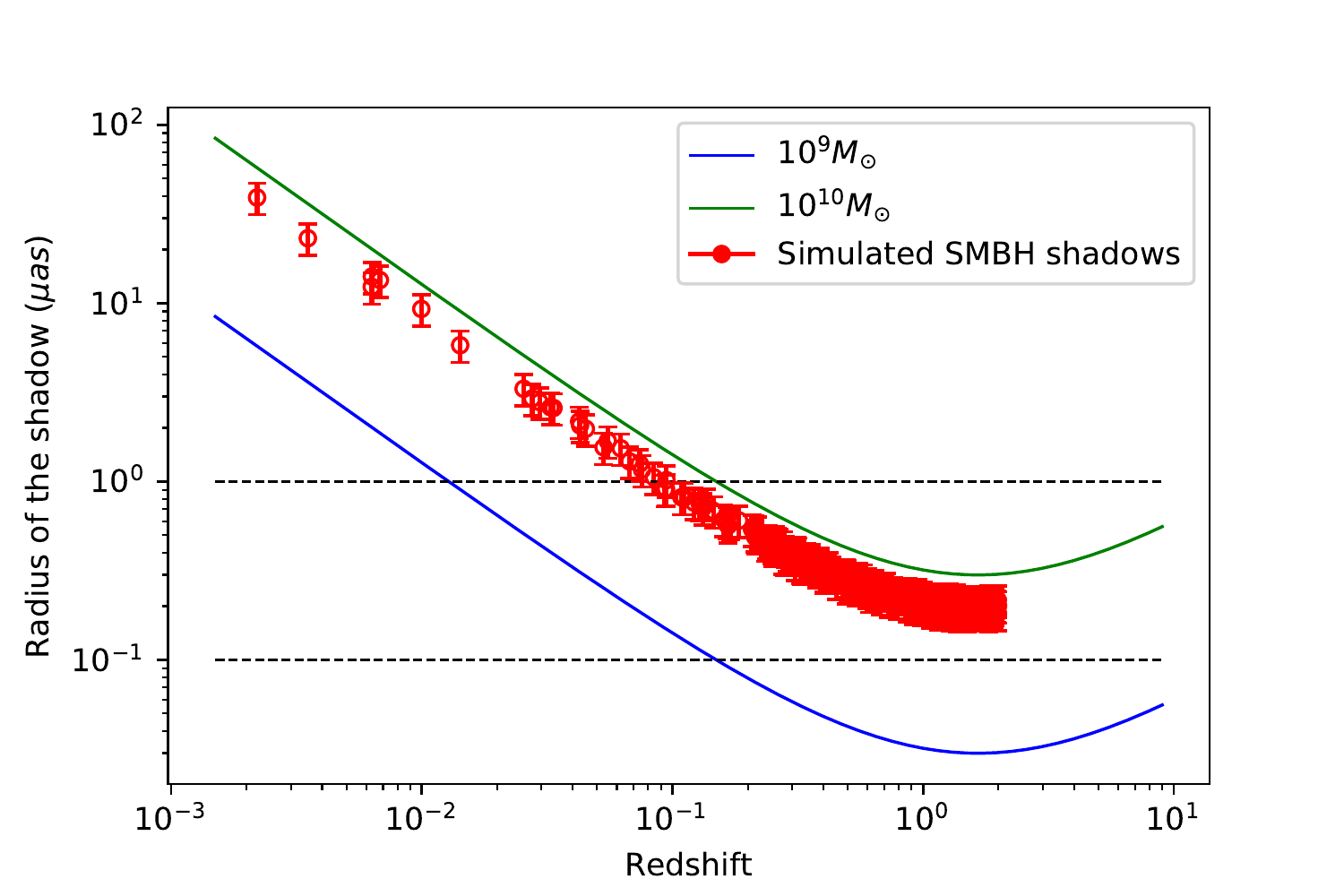}
    \caption{{\textbf{Left}}: {Simulations} for 10 SMBH shadows at high(er) redshift in $7 \leq z \leq 9$, with $10^9 - 10^{10} M_ \odot$. We consider a fiducial conservative cosmology with $H_0$ = 73.8 km/s/Mpc, $\Omega_{m0}=0.262$. Dashed lines indicate the values $\alpha_{sh}$ = 0.1 $\mu$as and $\alpha_{sh}$ = 1 $\mu$as, from bottom to top respectively.
    {\textbf{Right}}: Simulations for 500 SMBH shadows at low redshift in $0.0001 \leq z \leq 2.5$, with $5\times 10^9 M_\odot$. We consider the same latter cosmology setup. Dashed lines indicate the values $\alpha_{sh}$ = 0.1 $\mu$as and $\alpha_{sh}$ = 1 $\mu$as,  from bottom to top, respectively. 
    }
    \label{fig:simulationsBH}
\end{figure}

Analogous to our previous forecasting, we will now simulate the SMBH shadows at low redshift. For this case, as far as we know, most galaxies have a BH at their center; therefore, the possibility of observing them is greater in comparison to high(er) redshift ranges. We will simulate a conservative quantity of 500 SMBH shadows and generate random redshift values within $0 \leq z \leq 2.5$. We use a fixed BH mass within $5\times 10^ 6 M_\odot$. Our synthetic catalog consists of 500 random pairs $(z,M)$ from where we can obtain the radius of the BH shadow associated with each pair. In this case, we add a noise to our data that does not exceed our precision of $4\%$. Our forecasted data is shown at the right of Figure~\ref{fig:simulationsBH}. 

We will use both of these synthetic catalogs to implement Bayesian statistics in order to constrain the Hubble constant $H_0$.

\section{\label{sec:data}Current and Simulated Data Sets}

With the BH shadows forecasting now described, we can perform statistical analyses combining the simulated catalogs at low and high(er) redshifts with local observables as SNIa using a $\chi^2$-statistics method. The set of best fits $(h,\Omega_m)$ can be obtained through a process with a modified version on \texttt{emcee-PHOEBE}\footnote{\href{http://phoebe-project.org/docs/2.3/tutorials/emcee\_resampling}{phoebe-project.org}} for our cosmology and the new baseline (SNIa + BH shadows) and the extract of constraints using \texttt{GetDist}\footnote{\href{https://getdist.readthedocs.io/en/latest/}{getdist.readthedocs.io}}

In this paper we consider four different data sets:
\begin{itemize}
\item Pantheon SNIa catalog \cite{Pan-STARRS1:2017jku}. This catalog contains data of 1048 SNIa, observed in the range of low redshift from $0.01 < z < 2.3$. For each supernova, the redshift $z$ and the apparent magnitude $m(z)$ are given, which allows us to build the modulus distance $\mu$ by fixing the absolute magnitude $M$. In this analysis, we use the value $M = -19.3$, for one case. 

\item EHT direct observations. This set contains data from the two observations of the SMBH M87* and Sgr A*. For each observable, their mass $m$ is given in $M_\odot$ units, the redshift $z$ and  the radius of their shadow in $\mu$as units. A compilation of these data is given in Table \ref{tab:EHT data}.

\item High redshift SMBH shadows. This set contains 10 simulated shadows for SMBH between $7\leq z \leq 9$ (see Figure \ref{fig:simulationsBH} at the left). For each forecasted BH,  its redshift $z$, the size of its radius in $\mu$as and the error in this radius are given. Details are described in Section \ref{subsec:BHsimhigh}.

\item Low redshift SMBH shadows. This set contains 500 simulated shadows for SMBH between $0.1\leq z \leq 2$ (see Figure \ref{fig:simulationsBH} at the right).  For each forecasted BH  its redshift $z$, the size of its radius in $\mu$as and the error in this radius are given. Details are described in Section \ref{subsec:BHsimlow}.

\end{itemize}

\begin{table}[H] 
 \caption{Compilation of the BH events and their observations (data from \cite{EventHorizonTelescope:2019dse,EventHorizonTelescope:2022wkp}). The first column denotes the BH event and its reference; the second column, the $z$ at which they were observed; the third and fourth columns, the radius in microarc-second ($\mu$as) and the mass in solar masses units ($M_\odot$) of the BH event,~respectively. }
    \centering
\setlength{\tabcolsep}{6.26mm}
    \begin{tabular}{|c|c|c|c|} 
\toprule
\hline
         \textbf{Black Hole Event}& \textbf{Redshift}\boldmath{ $z$} & \textbf{Radius} \textbf{(\boldmath{$\mu$}as)} & \textbf{Mass} \textbf{(\boldmath{$M_\odot$})}  \\ 
         \hline \midrule
        M87*  & 0.00428 & 21$\pm$ 1.5 & $6.6 \pm 0.4 \times 10^9$\\ \midrule
        Sgr A* & 0.000001895 & 26.4 $\pm$ 1.15 & $4 \pm 0.32 \times 10^6 $ \\ \bottomrule
        \hline
    \end{tabular}
    \label{tab:EHT data}
\end{table}

Along with the analysis, we use the reduced Hubble constant $h$, defined as $h=H_0/100$ [km/s/Mpc]. In the case of observables reported in the Pantheon catalog, we employ:
\begin{equation}
    \chi^2_{\text{SNIa}} = \frac{1}{2} \sum^{N_\text{SNIa}}_{i=0}\frac{[\mu_{\text{SNIa}}-\mu_{th}(z;h,\Omega_m)]^2}{\sigma_{\text{SNIa}}^2},
\end{equation}
where $\mu_{\text{SNIa}}$ and $\sigma_{\text{SNIa}}$ denote the modulus distance and its error for the SNIa, and $\mu_{th}$ is the theoretical modulus distance, given by Equation~(\ref{eq:modulus}). The total sample consists of $N_\text{SNIa}=1048$ data points.
The $\chi^2$-statistics for the BH simulations can be expressed as:
\begin{equation}
    \chi^2_{\text{BH}} = \frac{1}{2} \sum^{N_{\text{BH}}}_{i=0} \frac{[\alpha_{obs}-\alpha_{th}(z; h, \Omega_m)]^2}{\sigma_{BH}^2},
\end{equation}
where $\alpha_{obs}$ is the observed radius of the shadow given by the EHT observations plus the SMBH shadow simulated at low and high $z$, $\alpha_{th}$ is the theoretical radius of the BH shadow given by Equation~(\ref{eq:luminosity shadow}) for high redshift simulations and by Equation~(\ref{eq:lowred}) for low redshift simulations and EHT observations, and $\sigma_{BH}$ are the errors in the observed/simulated radius from the BH shadow for each case. For EHT observations, $N_{\text{BH}}=2$; for high redshift simulations, $N_{\text{BH}}=10$ and for low redshift simulations $N_{\text{BH}}= 500$.
Our final statistical analysis consists of  the total baseline $ \chi^2_\text{T} = \chi^2_{\text{SNIa}} + \chi^2_{\text{BH}}$.


\section{\label{sec:results}Results}

\begin{figure}[H]
    \centering
    \includegraphics[width=9.0cm]{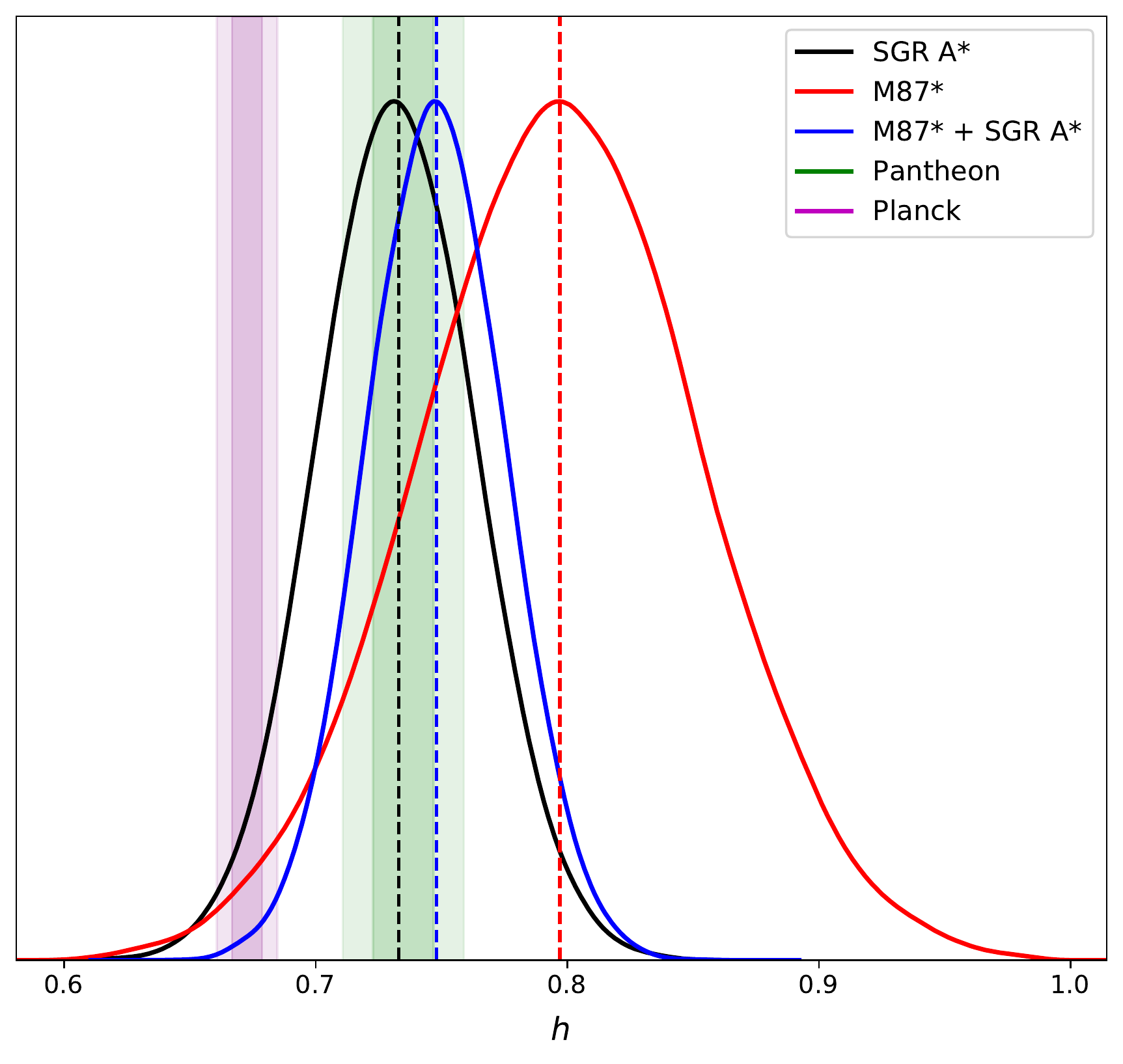}
     \caption{Reduced Hubble constant $h=H_0 / 100$ [km/s/Mpc] determined by our analysis using the SMBH shadows from M87* and Sgr A* combined observations at low redshift (blue color). Each individual posterior for M87* (red color) and Sgr A* (black color) are also showed. Vertical bands show the constraints at 1$\sigma$ for the Pantheon SNIa catalog (green color) and the Planck Collaboration (purple color). The vertical dashed lines indicate the mean value for $h$ .} 
    \label{fig:posteriorsSMBH}
\end{figure}

In Figure~\ref{fig:posteriorsSMBH}, we show the reduced $h$ {for}: {(i)} the SMBH shadows observed by the EHT array, and {(ii)} the combination of both SMBH shadows observed using our algorithm. Notice that the M87* shadow exceeds the SNIa Pantheon statistical range, which is obvious since we are computing a posterior with a single distant point in $z$ in comparison to the Sgr A* shadow (which is nearest to our Milky Way galaxy), whose $H_0$ values lies at 1$\sigma$ within the SNIa data set. Furthermore, the combination of M87* + Sgr A* gives a higher value of $H_0$ at the 1$\sigma$ border.

\begin{figure}[H]
\centering 
   \includegraphics[width=8.9cm]{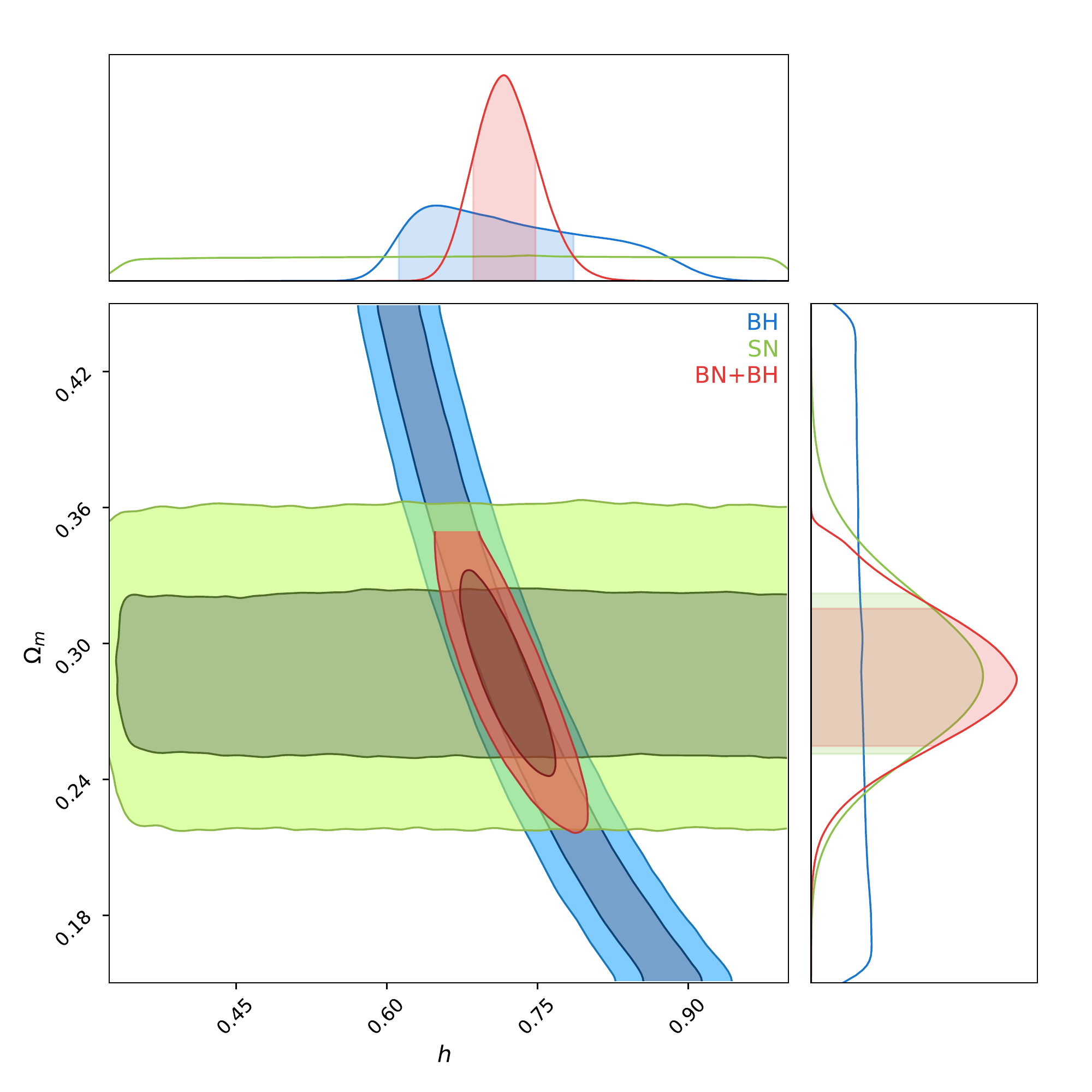}
        \includegraphics[width=8.9cm]{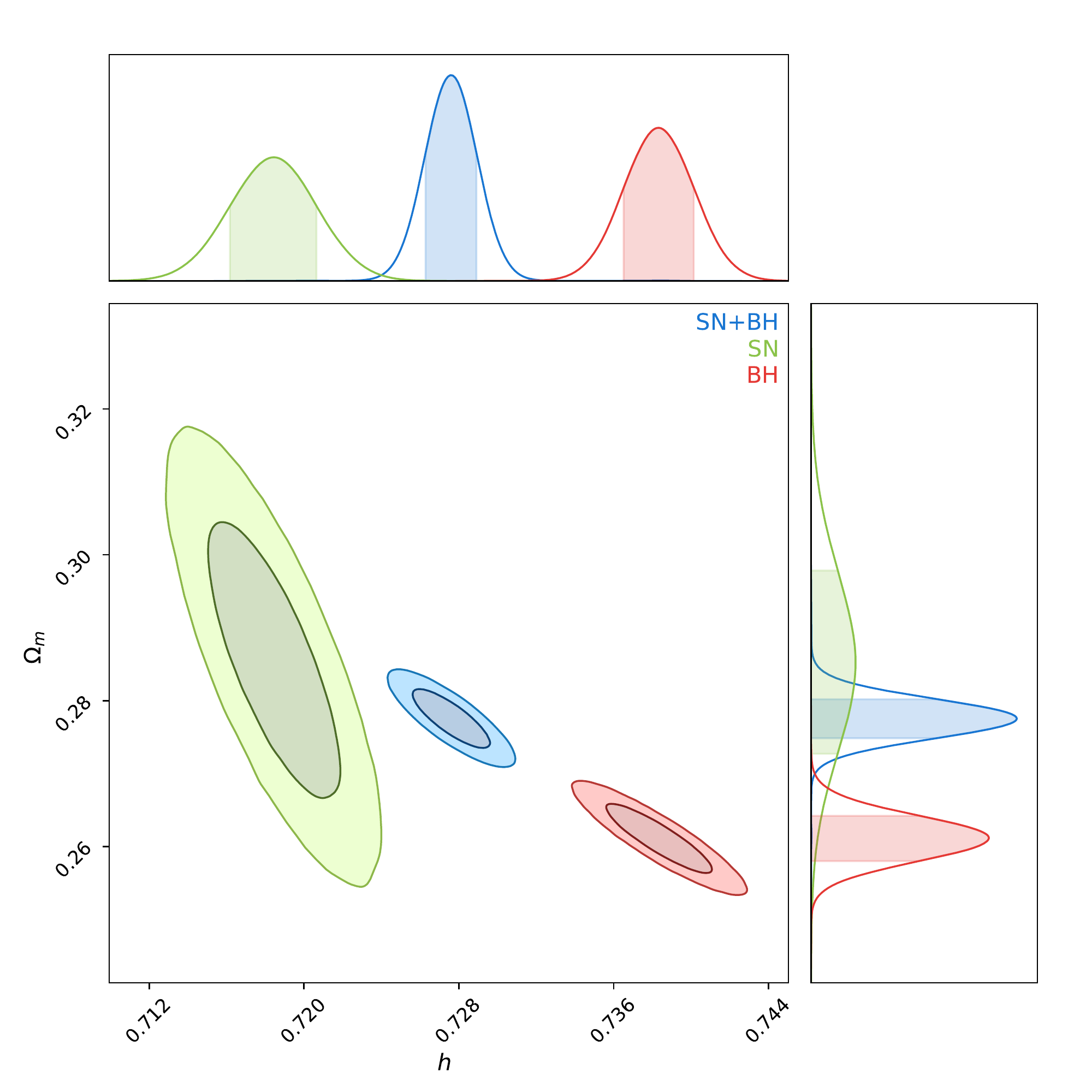}
    \caption{{\textbf{Left}:} Confidence contours C.L for the non-calibrated (full sample) Pantheon SNIa sample (here denoted by SN) and the high(er) redshift SMBH shadows simulations (here denoted by BH). Notice that SNIa (green color) gives a horizontal band that extends all over the range of values of $h$, which means that it can constrain the value for $\Omega_m$ but is not able to constrain $H_0$. The SMBH shadow simulations (blue color) form a region that extend widely along the values of $\Omega_m$. Our total statistics are given by the red C.L region, which allow usto constrain the cosmological parameters ($\Omega_m, H_0$), see Table \ref{tab:dataref}. {\textbf{Right}:} C.L for the calibrated (fixed $M=-19.3$) Pantheon SNIa sample and the low redshift SMBH shadows simulations (here denoted by BH). Notice that SNIa (green color) gives a wider contour over the range of values of $h$ and $\Omega_m$ in comparison to the SMBH shadows simulations (red color) that form a smaller region and give higher values for $h$. }
    \label{fig:calibration_notcalibration}
\end{figure}

In Figure~\ref{fig:calibration_notcalibration}, we show the confidence contours for each of our analyses: {(i)} When we consider the non-calibrated SNIa full catalog, a constraint value for $\Omega_m$ can be obtained, while the $H_0$ fails to be constrained. 
{(ii)} The simulated SMBH shadows at high(er) redshift have weak constraints on the $\Omega_m$ parameter. This leads to our final analysis: {(iii)} the combination of SNIa data plus SMBH mock data at low z, which can constrain the cosmological parameter in a local redshift range $(\Omega_m, H_0)$. {(iv)} Once we consider a calibrated SNIa full catalog (green color), we notice a tension between this catalog and the SMBH shadows simulations (red color). 

\begin{figure}[H]
    \centering
    \includegraphics[width=9.cm]{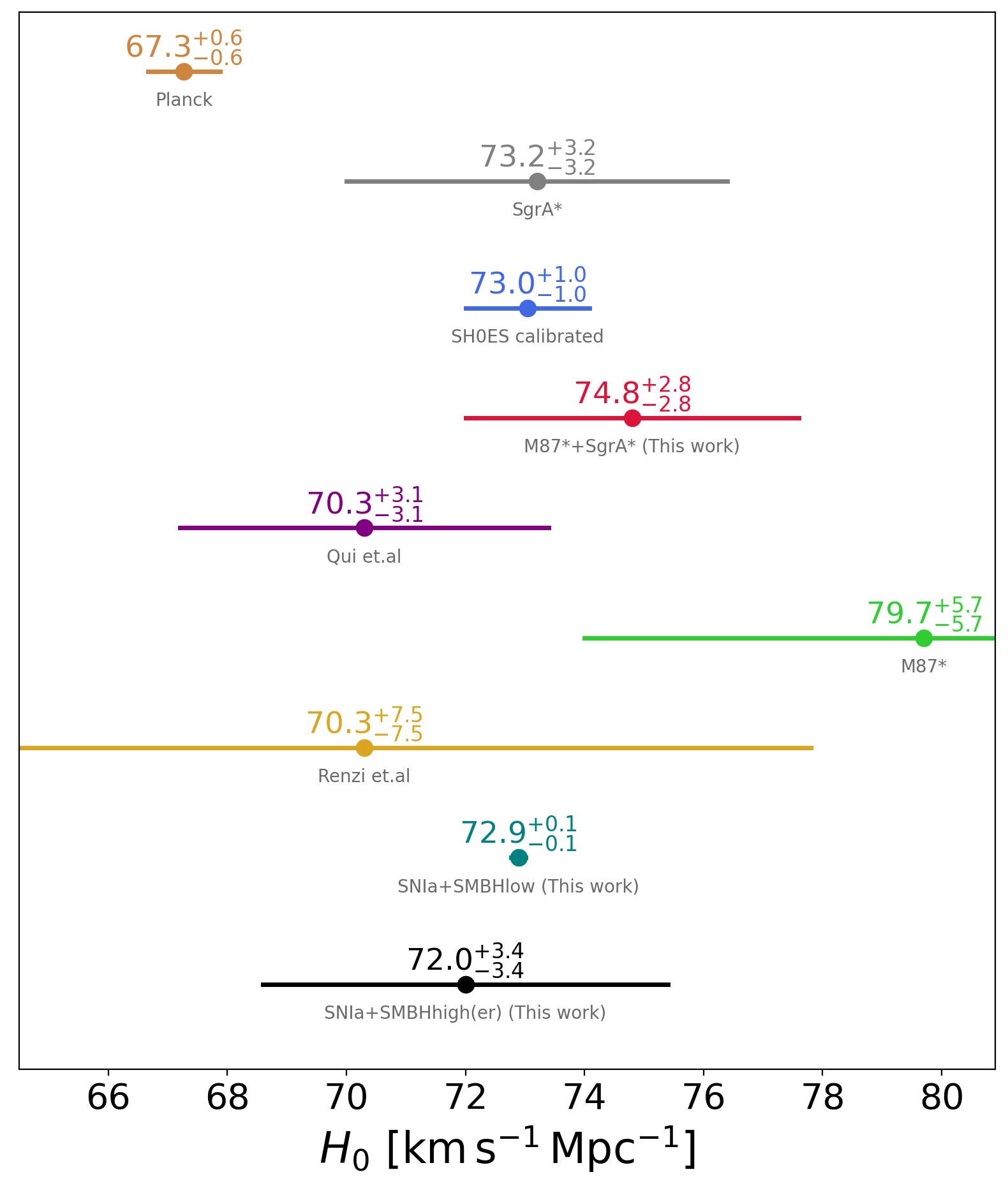}
    \caption{Whisker plot for the $H_0$ results reported in Table \ref{tab:dataref}.}
    \label{fig:H0whisker}
\end{figure}

A full compilation of the resulting pair $(\Omega_m, H_0)$ are given in Table \ref{tab:dataref} for each result reported in the literature, their baseline and the results obtained using our analyses. This table is complemented with a whisker plot given in Figure \ref{fig:H0whisker}. Notice that the value for $H_0$ using forecasting SMBH shadows at low $z$ under the assumptions described in our architecture gives an uncertainty lower than the ones reported in other methodologies. Additionally, our combined catalog SNIa + SMBH at high z reduces the tension in comparison to the direct EHT observations.

\vspace{-6pt}

\vspace{-6pt}

\begin{table}[H]
\caption{Compilations of results for ($H_0, \Omega_m$) constraints. The first column denotes the baseline and its references (data from \cite{Planck:2018vyg,Riess:2021jrx,EventHorizonTelescope:2019dse,EventHorizonTelescope:2022wkp,Qi:2019zdk,Renzi:2022fmw}, respectively). The second column indicates the observable/simulations treated in each baseline. The third and fourth columns indicate the ($H_0, \Omega_m$) values obtained from each analysis, respectively. Our results are indicated in the last two rows. In addition, we denote as {Fixed} when the baseline is considered flat prior to the parameter at hand, {Not reported} when the parameter was not computed,  and a dashed line (-) indicates a not related constraint since it was estimated at low redshift. \label{tab:dataref}}
\begin{adjustbox}{width=\textwidth}
\begin{tabular}	{| m{5cm}<{\centering} | m{5cm}<{\centering} | m{5cm}<{\centering} | c|} 
\toprule 
\hline
\centering \textbf{Base Line} & \centering \textbf{Observable/Simulations} & \centering \textbf{\boldmath{$H_0$} [km/s/Mpc]}& \boldmath{$\Omega_m$} \\ \midrule
\hline
\centering Planck collaboration  & \centering CMB & \centering  67.27 $\pm$ 0.604 & 0.315 $\pm$ 0.007 \\ 
\midrule
\hline
\centering SH0ES collaboration  & \centering Cepheid-SNIa sample with fixed $M=-19.253$ & \centering 73.04 $\pm$ 1.04 &  $0.297^{+0.023}_{-0.21}$ \\ 
\midrule
\hline
\centering EHT first observations & Size of  M87*  shadow. & \centering 79.7 $\pm$ 5.7 & - \\ \midrule \hline
\centering EHT second Observations  & Size of Sgr A* shadow . & \centering  73.2 $\pm$ 3.2 & - \\ \midrule \hline
\centering Both EHT Observations [This work]   & Sizes of M87* and Sgr A* shadows . & \centering  74.8 $\pm$ 2.8 & - \\ \midrule \hline
\centering Qi et al.& Size of    M87* shadow with a fixed mass using stellar-dynamics method plus SNIa from Pantheon catalog  & \centering  70.3 $\pm$ 3.1 & 0.301 $\pm$ 0.022 \\ 
\midrule \hline
\centering Renzi et al.  &  Size of    M87* shadow and mock catalogues for Supermassive BH ($\approx 10^6$ BH simulated) plus mock SNIa data for LSST. & \centering   $70.3 \pm 7.5$ 
& Fixed \\ 
\midrule \hline
\centering SNIa + SMBH at low redshift [This work] &  Sizes of    M87* and Sgr A* shadows, SNIa from Pantheon catalog plus forecasting for the sizes of SMBH shadows with \mbox{M = $3 \times 10^9 M_{\odot}$} at $z\le 0.01$ (see Section~\ref{subsec:BHsimlow}). & \centering  72.89  $\pm$ 0.12 & 0.275  $\pm$ 0.002 \\ 
\midrule \hline
\centering SNIa + SMBH at high redshift [This work]  & SNIa from Pantheon catalog plus forecasting for the size of the SMBH shadows with \mbox{M = $10^9 - 10^{10}~ M_\odot$} between \mbox{$7 \leq z \leq 9$} (see Section~\ref{subsec:BHsimhigh}). & \centering   72.0 $\pm$ 3.4 & 0.285 $\pm$ 0.029 \\
\bottomrule
\hline
\end{tabular}
 \end{adjustbox}
\end{table}


\section{\label{sec:discussion}Discussion}

In this paper, we proposed extending the studies of supermassive black holes shadows as standard rulers 
 order to study the $H_0$ tension. 
A BH cast a shadow in the neighborhood area of emission with a shape and size that can be computed using the location of the several photon orbits at different directions with respect to the spin axis. Furthermore, the angular size of shadows from a high redshift BH can increase due to cosmic expansion, hence the possibility to find constraints on the expansion history at high redshift.

The advantage of these SMBH shadows is the property that can be characterised by two parameters: the angular size of the shadow $\alpha$ at low and high redshifts and the BH mass. Moreover, a degeneracy between these parameters can arise at high redshift since assumptions on the precision of the experiment need to be taken into account. 

In order to break the degeneracy, in this paper we propose a viable forecasting method by fixing certain conditions at high(er) redshifts, i.e our results reach $\approx 4\%$, a precision that could be achievable in future experiments and with optimistic conditions. Furthermore, we found that our estimations provide a value of $H_0 = 72.89\pm 0.12$ and $\Omega_m=0.275\pm 0.002$, showing an improvement in the systematics 
reported so far in the literature for the SMBH standard rulers, including SN catalog in the total dataset analysis. Is important to mention that we recover the initial $H_0$ fixed prior (see Step 1 in Sec.\ref{sec:BHsim}) when solely it is considered the SMBH simulated sample (see the corresponding
result in red color in the right part of Figure \ref{fig:calibration_notcalibration}).
In addition, our value at high(er) redshifts, $H_0 = 72.0\pm 3.4$ and \mbox{$\Omega_m=0.285\pm 0.029$}, improves upon the systems reported at 8\% \cite{Renzi:2022fmw}, in fact, our systematic results are reduced to~4.7\%. 

We stress out that more general BH scenarios can be considered, for example, BH with spin and inclination variation as Kerr BHs. It was determined that these characteristics do not modify the determination of distances from the SMBH shadows \cite{Renzi:2022fmw}. In fact in general terms, the BH mass and the angular shadow size are the only two parameters that contribute significantly even in this spin-inclination BH system. 
Furthermore, we should mention that the treatment of a database that include SMBH forecastings and observational data like SN could bring relative weight issues combined with the fact that is necessary the assumption of an initial prior on $H_0$.
However, we expect that the precision obtained in this system can be improved using our methodology discussed here. We will report this elsewhere.

Currently, SMBH shadow observations are still  few; therefore the calculations 
derived from low data point statistics cannot accurately constrain  a set of cosmological parameters. However, as we presented in this paper, studying the precision assumptions that can be used for future experiments could allow us to promote these observables as future candidates in the many baselines used in cosmological tensions research.

\begin{acknowledgments}
The authors thank the referees and editor for some important comments which helped us to improve the paper considerably.
C.E.-R. is supported by the Royal Astronomical Society as FRAS 10147 and by PAPIIT UNAM Project TA100122. 
This article is based upon work from COST Action CA21136 Addressing observational tensions in cosmology with systematics and fundamental physics (CosmoVerse) supported by COST (European Cooperation in Science and Technology).

\end{acknowledgments}


\bibliographystyle{utphys}
\bibliography{references}

\end{document}